\documentclass[prd,showpacs,showkeys,floatfix,nofootinbib,fleqn,
tightenlines,eqsecnum,preprint]{revtex4}  

\usepackage{amsmath}\usepackage{amsfonts}
\usepackage{latexsym}\usepackage{graphicx}
\usepackage{mathrsfs}\usepackage{dcolumn}

\newcommand{\version}{v5}                

\newcommand{\beq}{\begin{equation}}
\newcommand{\eeq}{\end{equation}}
\newcommand{\beqa}{\begin{eqnarray}}
\newcommand{\eeqa}{\end{eqnarray}}
\newcommand{\bsubeqs}{\begin{subequations}}
\newcommand{\esubeqs}{\end{subequations}}
\newcommand{\half}{{\textstyle \frac{1}{2}}}  

\newcommand{\mass}{m}      
\newcommand{\gAbelian}{e}  

\begin{document}
%
%
%
\noindent arXiv:1202.0531  \hfill KA--TP--04--2012\;(\version)\vspace*{2\baselineskip}%
\title{Dynamically broken Lorentz invariance from the Higgs sector?}  

\author{F.R. Klinkhamer}
\email{frans.klinkhamer@kit.edu}
\affiliation{Institute for Theoretical Physics,
Karlsruhe Institute of Technology (KIT),\\ 76128 Karlsruhe, Germany\\}

\begin{abstract}
\vspace*{.125\baselineskip}\noindent  
The original Abelian $U(1)$ Higgs model in flat spacetime is
enlarged by the addition of one real scalar with a particular potential.
It is then shown that, while maintaining the original
masses of the vector boson and Higgs scalar,
there exists a time-dependent homogeneous
solution of the classical field equations, which
corresponds to dynamical breaking of Lorentz invariance (DBLI).
The same DBLI mechanism holds for the standard model
enlarged by the addition of a real isosinglet scalar with
an appropriate potential.
The resulting DBLI with an assumed TeV energy scale would manifest
itself primarily in the interactions of the two scalar particles.
In principle, this DBLI could feed into the neutrino
sector and give rise to a superluminal maximum velocity.
\vspace*{2\baselineskip}
\end{abstract}


\maketitle

\section{Introduction}\label{sec:intro}

The standard model (SM) of elementary particle physics is,
most likely, an effective
theory resulting from fundamental interactions at an energy scale
far above the electroweak energy scale (taken as
$E_\text{ew}=M_W \sim 10^2\;\text{GeV}$).

If this point of view is correct, it is possible
that the Higgs sector contains further scalar particles and interactions,
in addition to those of the minimal SM.
Here, we explore certain effects of an
additional real scalar field, which may lead to
dynamical 
breaking of Lorentz invariance
(DBLI).\footnote{\label{ftn:earlier-version-SBLI}In an earlier version
of this article, the
term `\textit{spontaneous} breaking of Lorentz invariance' (SBLI)
was used. But, for the moment,
the term `\textit{dynamical} breaking of Lorentz invariance'
is more appropriate.
Ultimately, the goal is to get rid of the global
(and gravitational) effects of
the condensate, with only flat Minkowski spacetime remaining and a
nonzero order parameter.
Then, it is perhaps also possible to get genuine
spontaneous symmetry breaking from the Higgs sector, independent
of the boundary conditions.}
This DBLI would manifest itself primarily in the Higgs sector
but perhaps also through nonstandard neutrino-propagation properties
(e.g., superluminal maximum velocities), as will be explained later on.

In order to keep the number of references manageable,
we only quote Ref.~\cite{Bjorken1963} for an early paper on
spontaneously broken Lorentz invariance,
Ref.~\cite{Higgs1966} for the Higgs mechanism,
and Ref.~\cite{ChengLi1984} for a textbook with the basics of
the SM and references to the original articles.
As far as the main idea of the present article is concerned,
DBLI from scalar fields,
we are not aware of a similar discussion in the literature
(Ref.~\cite{Anderson-etal2004}, for example, obtains bounds
on certain types of Lorentz violation in the Higgs sector,
but does not discuss the dynamic origin of the Lorentz violation).

It also needs to be emphasized
that the present article considers only toy-models for DBLI
by scalar fields,
leaving aside all questions about naturalness and renormalization.
The aim is really to see if it is at all possible to get
an acceptable form of DBLI
from the Higgs sector, more at the level of an existence proof
than a fully realistic theory.

\section{Abelian $\boldsymbol{U(1)}$ Higgs model}
\label{sec:Abelian-U1-Higgs-model}

\subsection{Nonstatic background solution}
\label{subsec:Nonstatic-background-solution-Abelian}

The starting point is the original Abelian $U(1)$ Higgs
model~\cite{Higgs1966,ChengLi1984} in terms of
the real gauge field $A_\alpha(x)$ and the complex scalar field
$\phi(x) \equiv [\phi_1(x) +i  \phi_2(x)]/\sqrt{2}$.
To this model is added a single real scalar $\xi(x)$ with
a particular potential. The Lagrange density
is taken as follows ($\hbar=c=1$):%
\bsubeqs\label{eq:L-Abelian}
\beqa\label{eq:L-Abelian-kinetic-minus-V}
\widehat{\mathcal{L}} &=&
-\frac{1}{4}\,F_{\alpha\beta}\,F^{\alpha\beta}
- \big( D_\alpha \phi \big)^{*}\,\big( D^\alpha \phi \big)
- \frac{1}{2}\,\big(\partial_\alpha \xi\big) \, \big(\partial^\alpha \xi\big)
-\widehat{V}\,,
\eeqa
\beqa
\label{eq:L-Abelian-V}
\widehat{V}&=&
\lambda\,  \big( |\phi|^2 -v^2/2 \big)^2
+\frac{1}{2}\,\mass^2\,\xi^2
+\zeta\,v^{-6}\,  \big( |\phi|^2 -v^2/2 \big)^4\,\xi^2\,,
\eeqa
\esubeqs
employing the usual Minkowski metric for the standard global spacetime
coordinates \mbox{$(x^\alpha)$ $=$ $(t,\,x^1,\,x^2,\,x^3)$,}%
\bsubeqs\label{eq:defs}
\beqa \label{eq:Minkowski-metric}
(\eta_{\alpha\beta})&=& \text{diag}(-1,\, 1,\, 1,\, 1)\,,
\eeqa
and the usual Maxwell field strength tensor and covariant derivative,
\beqa
F_{\alpha\beta}(x)&\equiv&
\partial_{\alpha} A_{\beta}(x)-\partial_{\beta}  A_{\alpha}(x) \,,\\[2mm]
D_\alpha \phi(x)  &\equiv&
\big[\partial_{\alpha} -i\, \gAbelian\, A_{\alpha}(x)\big]\,\phi(x)\,,
\eeqa
\esubeqs
where the nonzero real gauge coupling constant of this model is denoted
by $\gAbelian$ (no reference to the charge of the electron, just as
the vector field $A_\alpha$ has no reference to the photon field).

The theory \eqref{eq:L-Abelian} has
a local $U(1)$ gauge invariance involving the complex scalar
field, $\phi(x) \to \exp[i\,\gAbelian\, \omega(x)]\, \phi(x)$,
and a global $\mathbb{Z}_2$ invariance of the
real scalar field, $\xi(x) \to \pm\, \xi(x)$.
The parameters $v^2$, $\lambda$, $\mass^2$, and $\zeta$
in the potential \eqref{eq:L-Abelian-V} are real and bounded as follows:
\beqa\label{eq:param}
v^2 > 0\,,\quad \lambda> 0\,,\quad
\mass^2 > 0\,,\quad \zeta\geq 0\,.
\eeqa
The last term in the potential \eqref{eq:L-Abelian-V} is nonrenormalizable
and has mass-dimension 10. It is only added to provide an interaction
between the two scalar fields and can, in principle, be omitted
by setting $\zeta=0$. Incidentally, a single quartic coupling
$|\phi|^2\,\xi^2$ would not be compatible with the
generalized solution to be discussed shortly.

The classical field equations are:
\bsubeqs\label{eq:Abelian-field-eqs}
\beqa\label{eq:Abelian-field-eqs-A}
\partial^{\alpha}F_{\alpha\beta}
&=&
-2\, \gAbelian\;\,
\text{Im}\big(\phi^{*}\,D_{\beta}\phi \big)\,,
\\[2mm]
\label{eq:Abelian-field-eqs-phi}
D^\alpha D_\alpha\, \phi
&=&
2\lambda\, \big( |\phi|^2  -v^2/2 \big)\,\phi
+
4\,\zeta\,v^{-6}\, \big( |\phi|^2  -v^2/2 \big)^3\,\xi^2\,\phi\,,
\\[2mm]
\label{eq:Abelian-field-eq-xi}
\partial^\alpha \partial_\alpha\, \xi
&=&
\mass^2\,\xi+2\,\zeta\,v^{-6}\, \big( |\phi|^2  -v^2/2 \big)^4\,\xi\,.
\eeqa
\esubeqs
These classical field equations have
the standard Higgs vacuum solution~\cite{Higgs1966}:%
\bsubeqs\label{eq:Abelian-standard-Higgs-vac}
\beqa
A_{\alpha}(x)&=&0\,,
\\[2mm]
\phi(x) &=&v/\sqrt{2}\,,
\\[2mm]
\xi(x)&=& 0\,,
\eeqa
\esubeqs
up to a global phase of the field $\phi$.
This solution is static and homogenous:
the constant scalar fields correspond to a Lorentz-invariant state.

The background solution \eqref{eq:Abelian-standard-Higgs-vac}
can be generalized to a nonstatic homogenous solution,
without changing the
masses of the vector and scalar perturbation modes (see below).
Specifically, this generalized solution is
\bsubeqs\label{eq:Abelian-nonstandard-Higgs-vac}
\beqa\label{eq:Abelian-nonstandard-Higgs-vac-A}
A_{\alpha}(x)&=&0\,,
\\[2mm]\label{eq:Abelian-nonstandard-Higgs-vac-psi}
\phi(x) &=&v/\sqrt{2}\,,
\\[2mm]\label{eq:Abelian-nonstandard-Higgs-vac-xi}
\xi(x)&=& \xi_{0}\,\cos(\mass\,t) \,,
\eeqa
\esubeqs
for arbitrary real constant $\xi_{0}$.
The solution is only given up to a global phase of $\phi$
and up to time translation.

The nonstatic homogeneous background
\eqref{eq:Abelian-nonstandard-Higgs-vac}
with $\xi_{0} \ne 0$ can, in principle,
be selected by imposing appropriate initial boundary
conditions on the fields and their derivatives.
Remark that, for the case of vanishing coupling constant $\zeta$
and for the class of homogeneous field configurations,
\textit{generic} initial boundary conditions $\xi(t_\text{in})$ and
$\partial_{t}\xi(t_\text{in})$ give
a solution of the type \eqref{eq:Abelian-nonstandard-Higgs-vac-xi}
with $\xi_{0}\ne 0$, whereas only the special values
$\xi(t_\text{in})=0$ and $\partial_{t}\xi(t_\text{in})=0$ give a
solution with $\xi_{0}=0$.
What really needs to be explained is that these initial $\xi$ fields are
(approximately) homogeneous.
The explanation of the extraordinarily smooth conditions
just after the ``big bang'' is anyway
the major unsolved problem of modern cosmology
(the inflation mechanism may or may not play a role in
the final answer~\cite{Guth1981,Linde1983,Penrose2005}).
The present article, however, considers a fixed Minkowski
spacetime and neglects gravity altogether
(see also Sec.~\ref{subsec:DBLI-Abelian}).
For now, we simply assume that the initial boundary conditions
select solution \eqref{eq:Abelian-nonstandard-Higgs-vac}
with $\xi_{0}\ne 0$.

\subsection{Localized perturbations}
\label{subsec:Localized-perturbations-Abelian}
It remains for us to calculate the particle spectrum for the nonstatic
background \eqref{eq:Abelian-nonstandard-Higgs-vac}.
This can be done in unitary gauge~\cite{ChengLi1984},
\bsubeqs
\beqa
A_{\alpha}(x)&=&A_{\alpha}(x)\,,
\\[2mm]
\sqrt{2}\,\phi(x) &=& v+ h(x)\,,
\\[2mm]
\xi(x)&=&  \xi_{0}\,\cos(\mass\,t)+k(x) \,.
\eeqa
\esubeqs
Localized perturbations allow for partial integrations
without boundary terms
and the Lagrange density up to quadratic order is found to be given
by\footnote{\label{ftn:earlier-version-psi}In an earlier version
of this article, we considered
a theory with a complex (charged) scalar $\psi(x)$ instead of
the real (neutral) scalar $\xi(x)$ used here. However,
the quadratic Lagrange density given previously
missed two terms which may lead to instability.
For this reason, we now restrict
ourselves to having just an additional real scalar $\xi(x)$
and make sure to have only higher-order couplings between the
$h(x)$ and $k(x)$ modes.}
\bsubeqs\label{eq:linear-quadratic-interactions}
\beqa\label{eq:linear-quadratic}
\widehat{\mathcal{L}}^\text{\;(lin.+\,quadr.)}
&=&
-\frac{1}{4}\,(\partial_{\alpha} A_{\beta}-\partial_{\beta}  A_{\alpha})^2
-\frac{1}{2}\,\big[(\gAbelian\, v)^2\,\big]\,A_{\alpha}A^{\alpha}
\nonumber\\[1mm]
&&
-\frac{1}{2}\,(\partial_{\alpha} h)^2
-\frac{1}{2}\,\big[2\, \lambda\, v^2\,   \big]\,h^2
-\frac{1}{2}\,(\partial_{\alpha} k)^2
-\frac{1}{2}\,\big[ \mass^2  \big]\,k^2\,,
\eeqa
where, as expected, all linear terms have dropped out and a
zero-order time-dependent term has not been shown (this term
will be discussed in the last two paragraphs
of Sec.~\ref{subsec:DBLI-Abelian}).
The remainder of the Lagrange density includes
cubic and higher-order interaction terms,
\beqa\label{eq:interactions}
\widehat{\mathcal{L}}^\text{\;(interact.)}
&=&
-\gAbelian^{\;2}\,A^{\alpha}A_{\alpha}\,h\,(v+h/2)
- \lambda\,\left( v\,{h}^3 + {h}^4/4 \right)
\nonumber\\[1mm]
&&
-(\zeta/16)\,v^{-6}\,
\big( h^2 +2\,h\,v \big)^4\,\big( k+\xi_{0}\,\cos\mass\,t \big)^2
\,,
\eeqa
\esubeqs
where the $\zeta$ term contains monomials $h^a\,k^b$
for $a=4,\,\ldots \,,\, 8$ and $b=0,\, 1,\, 2$.
Interestingly, the background parameter $\xi_{0}$ only
affects the interactions of the scalar perturbation modes (see
Sec.~\ref{subsec:Nonstandard-Higgs-physics} for further discussion).

The scalar perturbations $h$ and $k$ in \eqref{eq:linear-quadratic}
have positive mass-squares, denoted by square brackets,
$(m_h)^2=2\, \lambda\, v^2>0$ and $(m_k)^2=m^2>0$.
The same holds for the vector-field perturbation,
$(m_A)^2=(\gAbelian\, v)^2>0$.
Hence, the classical stability of the nonstatic
background \eqref{eq:Abelian-nonstandard-Higgs-vac} is manifest
and independent of the background parameter $\xi_{0}$.
Quantum-dissipative effects from the radiation of $h$-quanta
can be made arbitrarily small by taking $\zeta$ sufficiently small.
For $\zeta=0$,
the nonstatic background \eqref{eq:Abelian-nonstandard-Higgs-vac}
is absolutely stable against localized perturbations.

\subsection{Dynamical breaking of Lorentz invariance}
\label{subsec:DBLI-Abelian}

It is possible to define the following tensor
composite of scalar-field derivatives:
\bsubeqs\label{eq:widehat-b-def-class-quant}
\beq\label{eq:widehat-b-def}
\widehat{b}_{\alpha\beta}\equiv
\frac{2}{\mass^{4}}\;
\big( \partial_\alpha\xi\big)\, \big(\partial_\beta\, \xi\big)\,,
\eeq
whose average over time intervals very much larger than $1/m$
does not vanish for the nonstatic classical
background \eqref{eq:Abelian-nonstandard-Higgs-vac},
\beq\label{eq:widehat-b-class}
\big<\widehat{b}_{\alpha\beta}\big>^\text{(nonstat.\,class.\,sol.)}_\text{time-average}
=
(\xi_{0}/\mass)^2\;
\delta_{\alpha,0}\,\delta_{\beta,0}\,.
\eeq
In the quantum theory with $\zeta=0$ (i.e., noninteracting $\xi$ fields),
the  corresponding time-averaged expectation value in a ground state
with dynamical symmetry breaking (DSB) is given by
\beq\label{eq:widehat-b-quant}
\widehat{b}_{\alpha\beta}\Big|^\text{(DSB,\,quant.\,th.)}_\text{time-average}
\equiv
\frac{2}{\mass^{4}}\;
\Big<
\big( \partial_\alpha\xi\big)\, \big(\partial_\beta\, \xi\big)
\Big>_\text{DSB,\;time-average}
\sim
(\xi_{0}/\mass)^2\;
\delta_{\alpha,0}\,\delta_{\beta,0}\,,
\eeq
\esubeqs
now in terms of the renormalized quantities
$\xi_{0}$ and $\mass^2$. Having a nonzero
order parameter \eqref{eq:widehat-b-class}
or \eqref{eq:widehat-b-quant} signals
the dynamical breaking of Lorentz invariance.

The composite operator \eqref{eq:widehat-b-def} used as a
diagnostic for broken Lorentz invariance is, of course, not unique.
A mathematically attractive alternative would be
\beq\label{eq:widetilde-b-def}
\widetilde{b}_{\alpha\beta}\equiv
\frac{\big( \partial_\alpha\xi\big)\, \big(\partial_\beta\, \xi\big)}
{\sqrt{\big[\big( \partial_\gamma\xi\big)\,
\big(\partial^\gamma\, \xi\big)\big]^2
+\epsilon\, m^8}}\;,
\eeq
for a positive infinitesimal $\epsilon$. The order parameter
$\widetilde{b}_{00}$ would be nonzero and approximately constant
also for $\xi(t)$ oscillations
with a slowly decreasing amplitude
(for example, due to quantum-dissipative effects if $\zeta$
is small but nonzero).
But, for the moment, we restrict the discussion to the simpler type of
composite \eqref{eq:widehat-b-def}.

Returning to the nonstatic background solution,
two technical remarks are in order. First,
given the fields equations \eqref{eq:Abelian-field-eqs},
there is no 
background solution equal
to \eqref{eq:Abelian-nonstandard-Higgs-vac}
with the argument $m t$ of the cosine function
replaced by, for example, $m x^1$. In this way, the
dynamics singles out the time components of the order
parameter \eqref{eq:widehat-b-class}.
Second, the background solution \eqref{eq:Abelian-nonstandard-Higgs-vac}
is invariant under time reversal (T),
charge conjugation (C), and parity reflection (P).
Hence, there is spontaneous breaking of Lorentz invariance
but not of CPT or any of the separate discrete symmetries.

Finally, there is an important issue, already alluded to in
Ftn.~\ref{ftn:earlier-version-SBLI}, which has to do with gravity.
In the standard approach,
solution \eqref{eq:Abelian-nonstandard-Higgs-vac} would give a
homogeneous pressure $P_{\phi,\,\xi}$
and energy density $\rho_{\phi,\,\xi}$, so that Minkowski spacetime
would no longer be a solution of the Einstein gravitational field
equations (see the Appendix for further discussion).
It is, of course, known that the Higgs mechanism has a
potential conflict with gravity, for example, as regards the
cosmological constant problem~\cite{Veltman1974,Weinberg1988},
and it is possible that the pressure and energy density
of the Higgs condensate gravitate unconventionally.
\mbox{\textit{A forteriori},} unconventional gravitational properties
may hold for the Lorentz-noninvariant state considered,
with $P_{\phi,\,\xi}+\rho_{\phi,\,\xi}\ne 0$.

For the moment, we choose to completely neglect gravity and to
postulate the flat Minkowski spacetime metric \eqref{eq:Minkowski-metric}.
Having fixed the metric \eqref{eq:Minkowski-metric},
the ground state \eqref{eq:Abelian-nonstandard-Higgs-vac}
is a perfectly stable classical solution, although with inherent
Lorentz breaking, as made clear by
the time-dependent interaction terms in \eqref{eq:interactions}
and the nonzero order parameter \eqref{eq:widehat-b-quant}.

\section{Enlarged standard model}
\label{sec:Enlarged-SM}

\subsection{Nonstatic background and DBLI}
\label{subsec:Nonstatic background solution-DBLI}

Now let us turn to the SM and use the more or less standard notation
of Ref.~\cite{ChengLi1984}. As the SM has only a single physical
scalar particle,
the previous DBLI mechanism cannot be implemented directly. It seems,
therefore, necessary to extend the scalar content of the SM.
One possibility is simply to add the two previous scalar fields
$\phi(x)$ and $\xi(x)$, without coupling to
the gauge field ($\gAbelian=0$) but keeping the previous
potential term \eqref{eq:L-Abelian-V}.
Then, however, there is no reason that the mass scales in
this potential would be of the order of the electroweak
scale $E_\text{ew}$.
Another possibility is to add only one neutral (sterile) scalar
field, now denoted $\Xi(x)$,
and to couple it via the potential term to the SM isodoublet $\Phi(x)$.

The Lagrange density is taken to be
\bsubeqs\label{eq:L-SM-extension}
\beqa\label{eq:L-SM-extension-action-dens}
{\mathcal{L}}&=&
\mathcal{L}_\text{SM,\,vector,\,spinor}
+\mathcal{L}_\text{scalar}\,,
\eeqa
with
\beqa\label{eq:L-SM-extension-action-dens-scalar}
\mathcal{L}_\text{scalar}
&=&
- \big( D_\alpha \Phi \big)^{\dagger}\,\big( D^\alpha \Phi \big)
- \frac{1}{2}\; \big(\partial_\alpha \Xi\big) \,   \big(\partial^\alpha \Xi\big)
-V_\text{scalar}\,,
\\[2mm]
\label{eq:L-SM-extension-action-dens-scalar-V}
V_\text{scalar}
&=&
\lambda\, \big( \Phi^{\dagger}\Phi  -v^2/2 \big)^2
+\frac{1}{2}\,\mass^2\:\Xi^2
+\zeta\,v^{-6}\,  \big( \Phi^{\dagger}\Phi -v^2/2 \big)^4\,\Xi^2
\,,
\eeqa
where $\Xi(x)$ is a real isosinglet and
$\Phi(x)$ a complex isodoublet with covariant derivative
\beqa
D_\alpha \Phi(x)  &\equiv&
\big[\partial_{\alpha}
+ g\, \tau^{a}/(2i)\,  W_\alpha^{a}(x)
+ g'\,\openone_{2}/(2i)  \,      B_{\alpha}(x)
\big]\, \Phi(x)\,,
\eeqa
\esubeqs
in terms of the $SU(2)$ gauge fields $W_\alpha^{a}(x)$
and the $U(1)$ gauge field $B_\alpha(x)$,
the three matrices $\tau^a$
being the usual $2\times 2$ Pauli matrices of isospin
and $\openone_{2}$ the $2\times 2$ unit matrix.
The parameters of potential
\eqref{eq:L-SM-extension-action-dens-scalar-V}
are taken as in \eqref{eq:param}.
More specifically, it may be natural to have
$v^2 \sim \mass^2$ and $\lambda\sim  \zeta \sim 1$,
with \eqref{eq:L-SM-extension} considered to be
an effective theory for energies up to the TeV scale.
Remark that, with a vanishing scalar
field $\Xi(x)\equiv 0$, the Lagrange density
${\mathcal{L}}$ in \eqref{eq:L-SM-extension-action-dens}
equals the standard-model Lagrange density~\cite{ChengLi1984},
$\mathcal{L}_\text{SM}$.

For later use, we explicitly give the $SU(2)$ representations
(isodoublet or isosinglet) of
the basic SM (anti-)fermion fields of the three lepton families
(label $f=e,\,\mu,\,\tau$) and the $SU(2)$ representation
(isodoublet) of the SM Higgs field:
\bsubeqs\label{eq:isodoublets-isosinglets}
\beqa\label{eq:isodoublets-isosinglets-leptons}
L_f(x) &=&
\left(
\begin{array}{c}
\nu_{f,L}(x) \\f^{-}_{L}(x) \\
\end{array}
\right)_{-1}
,\quad
R_{f}(x)  = \big(f^{+}_{R}(x)\big)_{+2} \,,\\[2mm]
\label{eq:isosinglets-Higgs}
\Phi(x) &=&
\left(
\begin{array}{c}
\phi^{+}(x)\\\phi^{0}(x) \\
\end{array}
\right)_{+1},\quad
\widetilde{\Phi} \equiv  i\tau_2\,\cdot\,\Phi^\ast
\equiv
\left(
  \begin{array}{cc}
    0 & \;+1 \\
    -1 & \;0 \\
  \end{array}
\right)
\cdot\,\Phi^\ast\,,
\eeqa
\esubeqs
where the suffix on the first three $SU(2)$ representations gives
the value of the $U(1)$ hypercharge $Y$ (recall that
the electric charge is given by $Q\equiv I_3+Y/2$, for
isospin $I_3$) and the asterisk in the last definition of
\eqref{eq:isosinglets-Higgs} denotes complex conjugation.
The isodoublet in \eqref{eq:isodoublets-isosinglets-leptons}
has lepton number $L=+1$ and the corresponding
isosinglet $L=-1$. As said, we also add the $Y=0$
isosinglet scalar $\Xi(x)$ with an appropriate potential
\eqref{eq:L-SM-extension-action-dens-scalar-V}.

Turning to the bosonic fields
of the enlarged-SM theory \eqref{eq:L-SM-extension},
the discussion is the same as for the Abelian $U(1)$ Higgs model
of Sec.~\ref{subsec:Nonstatic-background-solution-Abelian}.
Again, there is a nonstatic homogeneous Higgs-like solution,
\bsubeqs\label{eq:SM-nonstandard-Higgs-vac}
\beqa\label{eq:SM-nonstandard-Higgs-vac-W-B}
W_\alpha^{a}(x)&=& B_{\alpha}(x) = 0\,,
\\[2mm]
\label{eq:SM-nonstandard-Higgs-vac-Phi}
\Phi(x) &=&
v/\sqrt{2}\,\;
\left(
  \begin{array}{c}
    0 \\
    1 \\
  \end{array}
\right)\,,
\\[2mm]
\label{eq:SM-nonstandard-Higgs-vac-Xi}
\Xi(x)&=& \Xi_{0}\,\cos(\mass\, t) \,,
\eeqa
\esubeqs
up to global transformations of $\Phi$ and time translations.
Also, as mentioned in
Sec.~\ref{subsec:Nonstatic-background-solution-Abelian},
generic homogeneous boundary conditions favor having
\eqref{eq:SM-nonstandard-Higgs-vac-Xi} with $\Xi_{0} \ne 0$.
The discussion of the masses of the scalar and vector perturbation
modes is essentially the same
as in Sec.~\ref{subsec:Localized-perturbations-Abelian}.

It is, again, possible to identify a tensor composite of scalar fields,
\bsubeqs\label{eq:b-tensor}
\beq\label{eq:b-tensor-def}
{b}_{\alpha\beta}=
\frac{2}{\mass^{4}}\;
\big( \partial_\alpha\Xi\big)\, \big(\partial_\beta\, \Xi\big)\,,
\eeq
which has a nonzero time-averaged ${b}_{00}$ component
for the nonstatic background \eqref{eq:SM-nonstandard-Higgs-vac},
\beq\label{eq:b-tensor-value}
\big<
b_{\alpha\beta}\big>^\text{(nonstat.\,class.\,sol.)}_\text{time-average} =
\left(\frac{\Xi_{0}}{\mass}\right)^2\; \delta_{\alpha,0}\,\delta_{\beta,0}\,.
\eeq
\esubeqs
This background tensor will play an important role in
Sec.~\ref{sec:Scalar-neutrino-interactions}.

\subsection{Nonstandard Higgs physics}
\label{subsec:Nonstandard-Higgs-physics}

In our toy-model \eqref{eq:L-SM-extension} with
nonstatic background \eqref{eq:SM-nonstandard-Higgs-vac},
the two neutral scalar particles
(denoted $h$ and $k$, just as in Sec.~\ref{sec:Abelian-U1-Higgs-model})
have propagation properties and interactions which are described
by action-density terms similar to those in, respectively,
\eqref{eq:linear-quadratic} and \eqref{eq:interactions}.
The dispersion relations of these two scalars are entirely
standard (Lorentz-invariant),
only their interactions are nonstandard (i.e., spacetime-dependent).
In this subsection, it is assumed that the coupling constant $\zeta$
is nonzero and that the experiments run over time intervals during which
quantum-dissipative effects can be neglected.

The quartic coupling constant of the Higgs scalar $h$, for example,
is given by
\bsubeqs\label{eq:quartic-coupling-general-lowE}
\beq\label{eq:quartic-coupling-general}
\lambda_{h-\text{quart.}}=\lambda + \zeta^{\,\prime}\,\cos^{2}\mass\,t\,,
\eeq
with $\zeta^{\,\prime}\geq 0$ depending on the amplitude $\Xi_{0}$ of
the background solution, $\zeta^{\,\prime}\equiv (\Xi_{0}/v)^2\,\zeta$.
For low-energy processes ($\sqrt{s}\ll m \sim \text{TeV}$),
the effective coupling constant would be
\beq\label{eq:quartic-coupling-lowE}
\lambda_{h-\text{quart.}}^{\text{(low-energy)}}
\sim \lambda + \half\, \zeta^{\,\prime}\,.
\eeq
\esubeqs
The rapid oscillations in \eqref{eq:quartic-coupling-general}
would only show up for high-energy processes ($\sqrt{s}\sim  m$).
Concretely, the quartic Higgs coupling constant
determined from high-energy-collision experiments
(many identical experiments repeated over time)
would show an inherent uncertainty with spread $\zeta^{\,\prime}$,
independent of the experimental uncertainties involved.

Two parenthetical remarks are as follows. First,
the above discussion of the Higgs
self-coupling \eqref{eq:quartic-coupling-general-lowE}
refers to experiments in the preferred frame defined by
the background solution \eqref{eq:SM-nonstandard-Higgs-vac}.
In the laboratory frame,
there are (small) changes due to the
motion of the spinning Earth around the Sun and the motion of the
solar system as a whole with respect to the preferred frame
(see, e.g., Ref.~\cite{Bluhm-etal2003} for further discussion).
In addition to special-relativistic time-dilatation effects,
there may also be gravitational time-dilatation effects due to the
various solar-system masses. The proper analysis of these
gravitational time-dilatation effects
requires a complete solution of the combined field equations,
which lies outside the scope of this article as it neglects gravity
altogether (cf. the last paragraph of Sec.~\ref{subsec:DBLI-Abelian}).

Second, spacetime-dependent
coupling constants have been considered in other contexts
(see, e.g., Ref.~\cite{Uzan2010} for a review).
The focus, here, is on rapidly variable couplings
with ultrashort timescales, not cosmological timescales.
Another possible source of small-scale modulations of the
coupling constants may be a topologically nontrivial small-scale
spacetime structure~\cite{KlinkhamerRupp2003}.

All in all, the scalars $h$ and $k$
from \eqref{eq:linear-quadratic-interactions}
display a rather `mild' form of Lorentz violation or,
more precisely, Poincar\'{e} violation.
Still, this Lorentz/Poincar\'{e} violation with
spacetime-dependent coupling constants
has not been put in by hand but has arisen
dynamically.\footnote{\label{ftn:wild-DBLI}Remark that,
generally speaking, it does not make sense to
consider `wild' \textit{ad hoc} forms of Lorentz violation
if they cannot be generated dynamically.}
Theoretically, this is a significant improvement.
But important questions remain
[assuming the toy-model \eqref{eq:L-SM-extension}
to have any physical significance at all],
for example, the origin of
the special initial boundary conditions needed to select
the particular nonstatic homogeneous background solution
(cf. the last paragraph of
Sec.~\ref{subsec:Nonstatic-background-solution-Abelian}).

\section{Scalar-neutrino interactions}\label{sec:Scalar-neutrino-interactions}

\subsection{Higher-derivative term}
\label{subsec:Higher-derivative-term}

Let us discuss one further application
of our dynamic Lorentz-symmetry-breaking mechanism, namely,
as a model to describe the, as of yet unconfirmed, OPERA
result~\cite{OPERA2011} on a superluminal neutrino
velocity.\footnote{\label{ftn:CERN-press-release}%
A CERN press release (February 23, 2012)
from the OPERA Collaboration states that two possible sources of error
have been found and that new short-pulse measurements are scheduled
for May, 2012.}
It was soon realized~\cite{KlinkhamerVolovik2011,
Klinkhamer2011-flavor-relativity,Klinkhamer2011-superlum-sterile-nu,
Klinkhamer2011-neutrino-mass-SM}
that SBLI, from fermion condensates in particular,
could play a role in the explanation of the OPERA result.
Ref.~\cite{NojiriOdintsov2011} also mentions SBLI but really focusses
on DBLI for an explanation, although not giving a comprehensive dynamical
solution. With the background tensor \eqref{eq:b-tensor},
we can now propose a relatively simple DBLI model
which appears to be phenomenologically attractive (apart from one
outstanding problem as will be explained in
Sec.~\ref{subsec:Superluminal neutrinos}).

We use the enlarged SM of
Sec.~\ref{subsec:Nonstatic background solution-DBLI}
and, in particular, the fields \eqref{eq:isodoublets-isosinglets}
and the composite \eqref{eq:b-tensor-def}.
For now, we take for granted that the three neutrinos
obtain masses \mbox{$m_{\nu,\,n}\lesssim \text{eV}$,} for $n=1,2,3$.
We, then, introduce a further gauge-invariant interaction
term~\cite{Klinkhamer2011-neutrino-mass-SM} for the enlarged-SM fields:
\beqa\label{eq:L11}
\mathcal{L}_{11}(x)  &=&
\frac{2\,M}{v^2}\,\sum_{f}\,
\Big(\,\overline{L}_{f}(x)\cdot \widetilde{\Phi}(x)\,\Big)\,
\left[\,\frac{1}{M^2}\,
{b}^{\,\alpha\beta} \;\partial_{\alpha}\partial_{\beta}\,\right]
\Big(\, \widetilde{\Phi}^\dagger(x)\cdot L_{f}(x)\Big)^{c}
+ \text{H.c.} \,,
\eeqa
where $\psi^{c}(x)$ denotes the charge conjugate of field $\psi(x)$
and the prefactor $2/v^2$ is chosen to cancel the $\Phi$ contributions
from \eqref{eq:SM-nonstandard-Higgs-vac-Phi}.
The interaction term \eqref{eq:L11} is non-renormalizable
(with a suppression factor $1/M$ at low energies)
and violates lepton-number conservation.
Recalling the definition \eqref{eq:b-tensor-def} of
${b}_{\,\alpha\beta}$, it follows that
the composite field operator on the right-hand side of
\eqref{eq:L11} has mass dimension $11$, hence the suffix on the
left-hand side.
Without the insertion $\big[M^{-2}\;{b}^{\,\alpha\beta}\,
\partial_{\alpha}\partial_{\beta}\big]$
and replacing the prefactor $2M/v^2$ by $1/M_{5}$,
the resulting dimension-5
interaction term is precisely the Majorana-mass-type term
considered in Ref.~\cite{Weinberg1979} and, many years later,
in Ref.~\cite{Klinkhamer2011-neutrino-mass-SM}, where its
potential role for neutrino-LV was emphasized.

We assume all mass scales
entering \eqref{eq:L11} to be of the same order,
\bsubeqs\label{eq:ew-scales-widehat-b}
\beq\label{eq:ew-scales}
v \sim \mass \sim M \sim \text{TeV}\,.
\eeq
According to \eqref{eq:b-tensor}, the time-averaged
tensor ${b}^{\,\alpha\beta}$ in \eqref{eq:L11} is of order unity
for the nonstatic background \eqref{eq:SM-nonstandard-Higgs-vac}
with $\Xi_{0}\sim \mass$,
\beq\label{eq:ew-b}
{b}^{\,\alpha\beta}\sim \delta^{\alpha,\,0}\,\delta^{\beta,\,0}\,.
\eeq
\esubeqs
As mentioned in Sec.~\ref{subsec:DBLI-Abelian},
having a nonzero order parameter \eqref{eq:ew-b}
signals the dynamical breaking of Lorentz invariance.
In principle, it is also possible to use in
\eqref{eq:L11} the background tensor $\widetilde{b}^{\alpha\beta}$
from \eqref{eq:widetilde-b-def} with $\xi(x)$ replaced by $\Xi(x)$ .

\subsection{Superluminal neutrinos}
\label{subsec:Superluminal neutrinos}

In the nonstatic background \eqref{eq:SM-nonstandard-Higgs-vac}
with $\Xi_{0}\sim \mass$, the resulting Lorentz-violating interaction term
\eqref{eq:L11} leads to modified dispersion
relations of the neutrinos~\cite{KlinkhamerVolovik2011,
Klinkhamer2011-flavor-relativity,Klinkhamer2011-superlum-sterile-nu}.
For the three neutrino mass states ($n=1,\,2,\,3$) and
3-momenta $\mathbf{p}$ bounded
by $\max[(m_{\nu,\,n})^2] \ll |\mathbf{p}|^2 \ll \text{min}(M^2,\,m^2)$,
these dispersion relations are ($c=1$)
\beqa\label{eq:mod-disp-sbli-mass-basis}
\big(E_{\nu,\,n}(\mathbf{p})\big)^2
&\sim&
|\mathbf{p}|^2+\big(m_{\nu,\,n}\big)^2
+\big(\,{b}^{00}\,\big)^2\;M^{-2}\;|\mathbf{p}|^4\,,
\eeqa
\noindent
with ${b}^{00}\sim 1$ according to \eqref{eq:ew-b}.
Remark that the preferred frame of the Lorentz violation
in \eqref{eq:mod-disp-sbli-mass-basis} traces back to the
background solution \eqref{eq:SM-nonstandard-Higgs-vac}.
Most importantly, the quartic term in \eqref{eq:mod-disp-sbli-mass-basis}
is identical for all three neutrino mass states.
It is, of course, also possible to have a higher-order
Lorentz-violating term, for example,
a term proportional to $M^{-6}\,|\mathbf{p}|^8$ from the use of
two operator insertions with square brackets in \eqref{eq:L11}.
For large neutrino energies,
$|\mathbf{p}|^2 \gtrsim \text{min}(M^2,\,m^2)$,
the quartic momentum-dependence of the neutrino dispersion relations
\eqref{eq:mod-disp-sbli-mass-basis}
needs to be tempered, possibly by the introduction of further
higher-derivative terms~\cite{Klinkhamer2011-superlum-sterile-nu}.

Referring to the list of experimental ``facts''
given in Sec.~I of Ref.~\cite{Klinkhamer2011-superlum-sterile-nu}
(which contains further references in addition to those
given here), the situation is as follows:
\begin{enumerate}
\item[(i)]
The OPERA result~\cite{OPERA2011} $v/c- 1 \sim 10^{-5}$ for
the muon-neutrino time-of-flight velocity at an energy
of order $10\;\text{GeV}$,
assumed to be correct for the sake of argument,
can be explained by the modified dispersion relations
\eqref{eq:mod-disp-sbli-mass-basis}
if the mass parameters are of the electroweak scale
\eqref{eq:ew-scales-widehat-b},
possibly $\Xi_{0}\sim \mass \sim v$ and $M \sim 30\;\text{TeV}$
(it is already known that $v \sim 250\; \text{GeV}$).
Moreover, a narrow initial pulse of muon-neutrinos
at CERN would have negligible broadening after traveling
the $730$ km to the OPERA detector in the GranSasso Laboratory
(see the last paragraph of Sec.~3
in Ref.~\cite{Klinkhamer2011-superlum-sterile-nu}).
Incidentally, sterile-neutrino models in four or more spacetime
dimensions (see
Ref.~\cite{Klinkhamer2011-flavor-relativity}
and references therein) typically predict a substantial broadening
of the detected muon-neutrino pulse profile,
which is not what OPERA observes
(Sec.~9 of Ref.~\cite{OPERA2011}).
\item[(ii)]
The supernova SN1987a bound~\cite{Longo1987}
$|v/c- 1| \lesssim 10^{-9}$
on the electron-antineutrino velocity at an energy of order
$10\;\text{MeV}$
is satisfied because of the quadratic energy dependence of
the group velocity from \eqref{eq:mod-disp-sbli-mass-basis}.
\item[(iii)]
Coherent mass-difference neutrino oscillations
remain unaffected~\cite{Giudice-etal2011}, because the
Lorentz violation from \eqref{eq:mod-disp-sbli-mass-basis}
operates equally for all three neutrino masses
and, thereby, equally for all three neutrino flavors
[the original Lagrange density \eqref{eq:L11}
has indeed identical terms for all three families].
\item[(iv)]
Energy losses~\cite{CohenGlashow2011}
of the CERN--GranSasso neutrinos from the vacuum-Cherenkov
process $\nu_\mu\to \nu_\mu+Z^{0} \to \nu_\mu+e^{-}+e^{+}$
are significantly reduced~\cite{MohantyRao2011},
by a factor of approximately
$(3)^{-5/2}\sim 1/16$, compared to the losses in the
theory with an identical Lorentz-violating $|\mathbf{p}|^2$ term
in the three neutrino dispersion relations.
The heuristic argument~\cite{Klinkhamer2011-superlum-sterile-nu}
for the reduction factor $(1/\sqrt{3}\,)^{5}$ relies on the
effective-mass-square concept~\cite{KaufholdKlinkhamer2005}
applied to this muon-decay-type process, giving the vacuum-Cherenkov
rate $\Gamma \propto (G_F)^{2}\,(m_\text{eff})^{5}$.
\item[(v)]
The leakage of Lorentz violation from the
neutrino sector into the charged-lepton sector
by quantum effects appears to be
problematic~\cite{Giudice-etal2011}, especially
in view of the tight bounds on, e.g., the electron velocity.
Obviously, replacing $M\sim \text{TeV}$ in \eqref{eq:L11}
by a very much larger value such as
$M\sim 10^{10}\;\text{TeV}$ would reduce OPERA-like effects
from \eqref{eq:mod-disp-sbli-mass-basis} by a factor $10^{-20}$,
bringing the neutrino-sector Lorentz violation down to the level of
the current electron bounds, at least, for low enough energies.
\end{enumerate}

To conclude, it appears possible to have a scalar-DBLI model which
can describe OPERA's claimed result on a superluminal neutrino velocity
and other experimental facts of neutrino physics, see
items (i)--(iv) above. The difficulty is to connect to
experimental facts outside the neutrino sector, see item (v) above.
But it is precisely this difficulty which would make
the relatively large OPERA value for $v/c-1$,
if it turns out to be correct, so significant.

\section{Discussion}\label{Discussion}

In this article, we have shown (perhaps not for the first time)
that extended Higgs models in Minkowski spacetime can have
time-dependent homogeneous
solutions\footnote{\label{ftn:cond-mat}Possibly related phenomena have
been observed in condensed-matter physics, in particular,
Bose-Einstein-condensed states of coherent precession
in superfluid $^{3}\text{He}$; see
Ref.~\cite{BunkovVolovik2097} and further references therein.}
of the classical field equations, which
correspond to dynamical breaking of Lorentz invariance.
This holds, in particular, for a simple enlargement of the
standard-model Higgs sector with one extra real isosinglet scalar
and an appropriate potential. The energy scales of this potential
and the corresponding nonstatic background solution
are assumed to be at the TeV scale.

The dynamical breaking of Lorentz invariance
from a nonstatic scalar background may lead to
new effects in the Higgs sector such as
time-dependent couplings of the scalar particles.
In addition, this DBLI may feed into the neutrino
sector and give rise to superluminal maximum velocities,
with a velocity excess controlled by a mass scale $M\gtrsim \text{TeV}$.

At a more theoretical level, the fundamental problem is merging
this DBLI mechanism of scalar fields with gravity (see also the
Appendix). This must be done in such a way that the DBLI persists
over cosmological time scales and also meshes with the solution
of the cosmological constant problem. Incidentally,
the solution of the cosmological constant problem~\cite{Weinberg1988}
is still outstanding: there have been many suggestions
(for example, a dynamic adjustment
mechanism~\cite{Dolgov1997,KlinkhamerVolovik2008,EmelyanovKlinkhamer2011c})
but there is not yet a definitive solution.

In the present article, we have simply side-stepped the
problem of merging the scalar-DBLI mechanism and gravity
by considering only flat Minkowski spacetime.
This is sufficient for a preliminary investigation,
but, ultimately, gravity needs to be included.

\section*{\hspace*{-4.5mm}ACKNOWLEDGMENTS}\vspace*{-0mm}\noindent
G.E. Volovik is thanked for an interesting remark on condensed-matter
physics.

\begin{appendix}
\section{Flat-spacetime solution}
\label{app:Flat-spacetime-solution}

It has been remarked in the penultimate paragraph of
Sec.~\ref{subsec:DBLI-Abelian}
that the nonstatic scalar-field background
of the Abelian $U(1)$ Higgs model considered
does not allow for a Minkowski spacetime solution of the
Einstein equations, assuming a standard gravitational behavior
of the scalar condensate. In this appendix, we
present an example of how, in principle,
it may be possible to get a flat-spacetime solution
if further vacuum and matter contributions are included.

Consider, in fact, the introduction of two additional real scalar fields.
The first scalar field $\kappa(x)$ has the same kinetic and
mass terms as $\xi(x)$ in \eqref{eq:L-Abelian}
but no further interaction terms. The second scalar field $\chi(x)$ has
a ``wrong-sign'' kinetic term and no potential term at all
(scalars with a wrong-sign kinetic term have already been considered
in, for example, ghost-condensation models for infrared modifications
of gravity~\cite{ArkaniHamed-etal2003}).
In addition, there are fine-tuned initial
boundary conditions and a fine-tuned cosmological constant $\Lambda$.

Specifically, we add to the Lagrange
density $\widehat{\mathcal{L}}$  of \eqref{eq:L-Abelian-kinetic-minus-V}
the following four terms:
\beq\label{eq:hat-L-prime}
\widehat{\mathcal{L}}^{\;\prime}\big[A,\,\phi ,\,\xi,\,\kappa,\,\chi\big]
=
\widehat{\mathcal{L}}\big[A,\,\phi ,\,\xi\big]
-\half\,\partial_\alpha \kappa\, \partial^\alpha \kappa
-\half\,\mass^2 \kappa^2
+\half\,\partial_\alpha \chi\, \partial^\alpha \chi
-\Lambda\,.
\eeq
The field equation of $\kappa(x)$
is given by \eqref{eq:Abelian-field-eq-xi}
with $\xi(x)$ replaced by $\kappa(x)$ but without the $\zeta$
interaction term. The field equation of $\chi(x)$ is
$\partial_\alpha\partial^\alpha \chi=0$ and
the homogeneous solution takes the form $\chi(x)=c_1\, t+c_2$
with real dimensional constants $c_1$ and $c_2$.

Special boundary conditions are taken to select the following
homogeneous classical solution:
\bsubeqs\label{eq:app-solution}
\beqa\label{eq:app-solution-A}
A_{\alpha}(x)&=&0\,,
\\[2mm]
\label{eq:app-solution-phi}
\phi(x) &=&v/\sqrt{2}\,,
\\[2mm]
\label{eq:app-solution-xi}
\xi(x)&=& \mass\,\cos(\mass\,t) \,,
\\[2mm]
\label{eq:app-solution-zeta}
\kappa(x)&=& \mass\,\sin(\mass\,t) \,,
\\[2mm]
\label{eq:app-solution-chi}
\chi(x)&=&\pm\,\mass^2\, t+\chi_0\,,
\eeqa
with an arbitrary sign of the linear term in $\chi(t)$  and an
arbitrary additive constant $\chi_0$.
The above solution has an
equal amplitude for $\xi(t)$ and $\kappa(t)$,
chosen as $\xi_{0}=\mass$, and a nonzero phase difference,
chosen as $\pi/2$.\footnote{\label{ftn:theta}An alternative
formulation of the theory uses the complex (but neutral) scalar
field $\theta(x) = \xi(x)+i\,\kappa(x)$.
Then, the initial boundary conditions select
$\theta(x)=\mass\,\exp[i\,\mass\,t]$.}
Moreover, the cosmological constant is fine-tuned
to the following negative value:
\beq\label{eq:app-solution-Lambda-value}
\Lambda=-\half\, \mass^4\,.
\eeq
\esubeqs
It is also possible that an effective cosmological constant with the
precise value \eqref{eq:app-solution-Lambda-value} arises dynamically
without
fine-tuning~\cite{Dolgov1997,KlinkhamerVolovik2008,EmelyanovKlinkhamer2011c}.
But, here, we simply postulate the appropriate cosmological
constant \eqref{eq:app-solution-Lambda-value}.
Observe that, with metric signature \eqref{eq:Minkowski-metric},
the contributions of the solutions \eqref{eq:app-solution-chi}
and \eqref{eq:app-solution-Lambda-value} cancel in
the action density \eqref{eq:hat-L-prime}.

The homogeneous background fields
\eqref{eq:app-solution-phi}, \eqref{eq:app-solution-xi}, and
\eqref{eq:app-solution-zeta} give the following contributions to the
pressure and the energy density:%
\bsubeqs\label{eq:P-rho-from-phi-xi-zeta}
\beqa
P_{\phi,\,\xi,\,\kappa}&=&
|\dot{\phi}|^2+\half\,(\dot{\xi})^2+\half\,(\dot{\kappa})^2
-\widehat{V}(\phi,\,\xi)-\half\,\mass^2\,\kappa^2=0\,,
\\[2mm]
\rho_{\phi,\,\xi,\,\kappa}&=&
|\dot{\phi}|^2+\half\,(\dot{\xi})^2+\half\,(\dot{\kappa})^2
+\widehat{V}(\phi,\,\xi)+\half\,\mass^2\,\kappa^2= \mass^4\,,
\eeqa
\esubeqs
where
the overdot stands for differentiation with respect to the coordinate
time $t$. The wrong-sign scalar field \eqref{eq:app-solution-chi}
contributes%
\bsubeqs\label{eq:P-rho-from-chi}
\beqa
P_{\chi}    &=& -\half\, (\dot{\chi})^2=-\half\, \mass^4\,,
\\[2mm]
\rho_{\chi} &=& -\half\, (\dot{\chi})^2=-\half\, \mass^4\,,
\eeqa
\esubeqs
and the cosmological constant \eqref{eq:app-solution-Lambda-value} gives
\bsubeqs\label{eq:P-rho-from-Lambda}
\beqa
P_{\Lambda}    &=& -\Lambda=+\half\, \mass^4\,,
\\[2mm]
\rho_{\Lambda} &=& +\Lambda=-\half\, \mass^4\,.
\eeqa
\esubeqs
With vanishing gauge-field background \eqref{eq:app-solution-A},
the total pressure and energy-density are nullified,
\bsubeqs\label{eq:P-rho-total}
\beqa\label{eq:P-total}
P_\text{total}^\text{(background)}
&=& P_{\phi,\,\xi,\,\kappa}+P_{\chi} +P_{\Lambda}=0\,,
\\[2mm]
\label{eq:rho-total}
\rho_\text{total}^\text{(background)}
&=& \rho_{\phi,\,\xi,\,\kappa}+\rho_{\chi} +\rho_{\Lambda}=0\,.
\eeqa
\esubeqs
Having only background fields contributing (i.e., no excitations),
flat Minkowski spacetime is then a solution of the Einstein
equations. Indeed, the Hubble parameter $H\equiv \dot{a}/a=0$ of
Minkowski spacetime solves the spatially-flat Friedmann
equations $H^2 =(8\pi/3)\, G_{N}\,\rho_\text{total}$
and $2\dot{H}+3H^2 =8\pi\, G_{N}\,P_\text{total}$.
But, even in a fictitious world without gravity ($G_N=0$),
the pressure condition \eqref{eq:P-total} is needed to describe
a self-sustained equilibrium state
(cf. Secs. II A and IV C in Ref.~\cite{KlinkhamerVolovik2008}).

Classically and with localized perturbations of the
original $A(x)$, $\phi(x)$, and $\xi(x)$ fields,
it is possible, in first approximation,
to neglect the existence of the additional fields $\kappa(x)$ and
$\chi(x)$, as these fields have no interactions
with the other fields (apart from gravitational interactions).
The background matter fields \eqref{eq:app-solution-zeta}
and \eqref{eq:app-solution-chi}
ensure having a constant background pressure and energy density,
and do not play a role in the local physics,
as long as localized gravitational interactions can be neglected
(or in the fictitious world without gravity, \mbox{$G_N=0$}).

Admittedly, the example of this appendix is completely
\textit{ad hoc} and physically unconvincing.
But the example does demonstrate
that, in principle, it may be possible to recover the
flat-spacetime solution of the standard Einstein equations
even if there is a nonstatic scalar
background \eqref{eq:Abelian-nonstandard-Higgs-vac}.

\end{appendix}


\end{document}